\documentclass[twoside,twocolumn,english,nofootinbib]{revtex4}
\usepackage[T1]{fontenc}
\usepackage[utf8]{inputenc}
\usepackage{amssymb}
\usepackage{esint}

\makeatletter

\usepackage{babel}
\makeatother

\begin{document}

\title{Stationary Configurations Imply Shift Symmetry: No Bondi Accretion
for Quintessence / k-Essence.}

\author{Ratindranath Akhoury }

\email{akhoury@umich.edu}

\author{Christopher S. Gauthier}

\email{csg@umich.edu}

\affiliation{Michigan Center for Theoretical Physics,\\
 Randall Laboratory of Physics,\\
 University of Michigan, \\
Ann Arbor, MI 48109-1120, USA}

\author{Alexander Vikman}

\email{alexander.vikman@nyu.edu}

\affiliation{Center for Cosmology and Particle Physics, \\
Department of Physics, New York University,\\
4 Washington Place New York, NY 10003, USA}

\date{\today}

\begin{abstract}
In this paper we show that, for general scalar fields, stationary
configurations are possible for shift symmetric theories only. This
symmetry with respect to constant translations in field space
should either be manifest in the original field variables or reveal itself
after an appropriate field redefinition. In particular this result
implies that neither k-\emph{Essence} nor \emph{Quintessence} can
have exact steady state / Bondi accretion onto Black Holes. We also
discuss the role of field redefinitions in k-\emph{Essence }theories.
Here we study the transformation properties of observables and other
variables in k-\emph{Essence} and emphasize which of them are covariant
under field redefinitions. Finally we find that stationary field configurations
are necessarily linear in Killing time, provided that shift symmetry
is realized in terms of these field variables. 
\end{abstract}


\maketitle

\section{Introduction}

The discovery of the late time acceleration of the universe using
Supernova Ia \cite{Perlmutter:1997zf,Riess:1998cb} confirmed by
other observations (see Ref. \cite{Bahcall:1999xn} and references therein),
opened a window of opportunity for the existence of novel cosmological
scalar fields not only during the early inflationary stage but also
in the current universe. Indeed the scalar fields are the most natural
candidates for realization of inflation and for the dynamical explanation
of \emph{Dark Energy} (DE) which is responsible for the late time
acceleration. Arguably, the main difficulty in the modeling and understanding
of the possible dynamics of \emph{Dark Energy}, arises because of
the fine tuning issues. In particular, there is the so-called
coincidence problem \cite{Steinhardt97,Zlatev:1998tr}: why the energy
density in DE is only now comparable with the energy density in the
dust-like \emph{Dark Matter? }This coincidence would be especially
remarkable, if one assumes that both these Dark constituents
are independent of each other and evolve very differently in time.
Partially because of the fine tuning problems it is not surprising
that the candidates for DE often have not only rather exotic names:
\emph{Quintessence / Cosmon } \cite{Wetterich:1987fm,Ratra:1987rm,Peebles:1987ek,Caldwell:1997ii,Zlatev:1998tr,Steinhardt:1999nw}\emph{,}
k-\emph{Essence} \cite{ArmendarizPicon:2000ah,ArmendarizPicon:2000dh,Chiba:1999ka},
\emph{Phantom} \cite{Caldwell:1999ew}, \emph{Ghost
Condensate} \cite{ArkaniHamed:2003uy}, \emph{Quintom} \cite{Feng:2004ad}
etc but also correspondingly very unusual properties. In particular,
these scalar fields can possess: extremely small effective mass (\emph{Quintessence},
\emph{Quintom}), sound speed which can be much smaller and even larger
than the speed of light (k-\emph{Essence}, \emph{Ghost Condensate}),
negative kinetic energies (\emph{Phantom}, \emph{Quintom}), Lorentz
symmetry breaking and gravity modifications even around the Minkowski
space-time background (\emph{Ghost Condensate}). The most successful
paradigm to solve the coincidence problem is currently the k-\emph{Essence,}
where\emph{ }the highly nonlinear dynamics triggers the equation of
state of DE from radiation-like to quasi de Sitter around the transition
to the matter domination stage. In the late matter domination epoch
the k-\emph{Essence} has the speed of sound which is much smaller
than one. However, it was showed \cite{Bonvin:2006vc,Kang:2007vs}
that to explain the coincidence problem k-\emph{Essence} models
must necessarily have, at least, a short phase where the fluctuations in the k-\emph{Essence} travel at superluminal speeds. For our paper it is important that the nonlinear
dynamics responsible for the attractor behavior addressing the coincidence
problem requires an explicit dependence of the Lagrangian on the field
strength \cite{Kang:2007vs}. This field dependence cannot be eliminated
by any field redefinitions. Thus, successful k-\emph{Essence }models
as well as \emph{Quintessence / Cosmon} models cannot be shift symmetric. 

On the other hand it is known that the current universe is highly
inhomogeneous on small scales and in particular that there are plenty
of Black Holes (BHs) of different mass and origin. Thus an interesting
and natural question arises, how do Black Holes surrounded by
cosmological scalar fields evolve? In addition, from the theoretical
viewpoint it is interesting to consider BHs {}``dressed'' with different
field backgrounds. This could have a valuable impact on our understanding
of the physics of horizons (see e.g. \cite{Dubovsky:2006vk,Babichev:2006vx,Babichev:2007wg,Dubovsky:2007zi}).
Owing to the no-hair theorems \cite{Bekenstein:1996pn,Mayo:1996mv,Sudarsky:1995zg,Heusler:1994wa,Heusler:1992ss}
we know that BHs cannot support static configurations of scalar
fields %
\footnote{BHs can not support scalar hair at least for theories that respect
some of the standard energy conditions. Having in mind the exotic properties of DE
models mentioned above, it would be interesting to find examples of
stable scalar hair in theories violating the usual energy conditions.
For a model of hairy scalar BHs with ghost like quantum instabilities
see Ref. \cite{Bronnikov:2005gm}. %
}. Therefore, any scalar hair will be continuously swallowed by the BH.
In particular one could analyze the growth (and may be even formation)
of Black Holes due to the accretion (collapse) of DE. Then one can
try to use powerful and rather universal laws of Black Hole thermodynamics
combined with astrophysical observations to restrict the allowed properties
of DE candidates and rule out some of them as contradicting to
either BH thermodynamics or astrophysical data. Recent studies along
these lines were, for example, done e.g. in Ref. \cite{Babichev:2004yx,Frolov:2002va,Frolov:2004vm,Babichev:2005py,Babichev:2006vx,Babichev:2007wg,Babichev:2008dy,Babichev:2008jb,Mukohyama:2005rw,Bean:2002kx,Dubovsky:2007zi,Dubovsky:2006vk}. 

Finally, for k-\emph{Essence,} a typically very small sound speed during
the late matter domination era allows for rather significant large-scale
inhomogeneities around BHs and other massive objects. This long-range
clumping would be one of the characteristic, potentially observable
consequences of k-\emph{Essence}. Moreover, due to this ability to
realize small sound speeds along with the dust-like equation of state,
the k-\emph{Essence} fields can be used to model \emph{Dark Matter} \cite{ArmendarizPicon:2005nz,Bertacca:2007fc,Akhoury:2008dv}. In
this setup, the presence of supermassive BHs at the center of galaxies
makes understanding the accretion process even more necessary. 

On the other hand the presence of backgrounds with the superluminal
sound speed mentioned above opens an exciting possibility to look
beyond the BH horizon \cite{Babichev:2007wg,Babichev:2006vx}%
\footnote{Despite of the presence of the superluminal propagation the accretion
backgrounds constructed in these works are free of any causal pathologies \cite{Babichev:2007dw}. However, it is interesting to study whether,
similar to Ref. \cite{Dolgov:1998fp}, two boosted BH could create
causal paradoxes in this setup. %
}. Note that the current bounds \cite{Erickson:2001bq,Hannestad:2005ak,TorresRodriguez:2007mk,TorresRodriguez:2008et,Xia:2007km}
on DE sound speed are not restrictive at all. 

The classical and most simple setup for accretion problems
is a steady state or Bondi accretion \cite{1952MNRAS.112..195B}.
Remarkably, a lot of astrophysical phenomena can be described by a steady
state accretion. For a review see e.g. \cite{Beskin:2002si}. For
scalar fields, the Bondi accretion was recently studied in \cite{Jacobson:1999vr,Frolov:2002va,Frolov:2004vm,Babichev:2008dy,Babichev:2007wg,Babichev:2006vx}.
It is fair to say that almost all known analytical solutions %
\footnote{See however Ref. \cite{Bean:2002kx} %
} for accreting scalars either belong to the Bondi case or represent
the dust-like free fall. The dust-like time dependent accretion of
a massive canonical scalar field was considered in \cite{Frolov:2002va},
while\emph{ } dust-like solutions for the\emph{ Ghost Condensate} scalars were found in \cite{Mukohyama:2005rw,Dubovsky:2007zi}.
It seems that scalar fields with canonical kinetic terms would not
leave any important impact on the astrophysical BHs in the current
universe \cite{Frolov:2002va}. Nevertheless, accreting scalar fields
could play an important role for the formation of primordial BHs (see
e.g. \cite{Bean:2002kx}). 

In this paper we investigate stationary configurations for general
k-\emph{Essence} scalar field theories. We show that the necessary
condition for the existence of exact stationary configurations is
the symmetry of the theory with respect to constant shifts in
the field space: $\phi\rightarrow\phi+c$. This symmetry has to be
realized either in terms of the original field strength or after a
field redefinition. On the way, we also analyze properties of general
k-\emph{Essence} scalar field theories covariant with respect to field
redefinitions. The proof is valid for general theories with nonlinear
kinetic terms in both the test-field approximation and the self-consistent
case where the background metric is governed by the field $\phi$
itself. It is interesting to note that shift symmetric scalar field
theories are exactly equivalent to perfect fluid hydrodynamics provided
that only such field configurations which have timelike
derivatives are considered. In particular this result implies that the most interesting
scalar field models of \emph{Dark Energy} cannot realize a steady
state / Bondi accretion. Thus, in general, the solution to the problem
of accretion of these fields onto Black Holes requires a knowledge
of their initial configuration. In this paper we are discussing stationary
configurations, which are exact. Of course for the real world the
stationarity should be considered as an approximation. It may well
happen that the solutions would only asymptotically approach the stationary
regime. For some canonical scalar fields this behavior was demonstrated
in \cite{Frolov:2002va}.

\section{Derivation of the stationary configurations}
Let us consider a general scalar field theory with the action 
\begin{eqnarray*}
S=\int\mbox{d}^{4}x\sqrt{-g}P\left(\phi,X\right), & \mbox{where} & X=\frac{1}{2}g^{\mu\nu}\nabla_{\mu}\phi\nabla_{\nu}\phi,
\end{eqnarray*}
$g_{\mu\nu}$ is the gravitational metric and as usual $g\equiv\mbox{det}g_{\mu\nu}$.
Throughout the paper $\nabla_{\mu}$ is the covariant derivative associated
with the gravitational metric $g_{\mu\nu}$. We assume that the
Lagrangian $P\left(\phi,X\right)$ is a general function satisfying
the following conditions: $P_{,X}\geq0$ (Null Energy Condition) and
$2XP_{,XX}/P_{,X}>-1$ (Hyperbolicity condition) %
\footnote{In this paper we use the notation $(...)_{,X}\equiv\partial\left(...\right)/\partial X$
and the signature $\left(+---\right)$. %
}. The first condition guaranties that the perturbations carry positive
kinetic energy while the second one implies the stability with respect
to high frequency perturbations and is necessary for the Cauchy problem
to be well posed (see e.g. Refs. \cite{Aharonov:1969vu,ArmendarizPicon:2005nz,Rendall:2005fv,Babichev:2007dw,Bruneton:2006gf,Bruneton:2007si}).
These conditions restrict the variety of the allowed Lagrangians along
with the corresponding solutions and are unavoidable for any physically
meaningful model %
\footnote{For a different opinion see \cite{Creminelli:2008wc}.%
}. The energy-momentum tensor of the theory is 
\begin{equation}
T_{\mu\nu}=P_{,X}\nabla_{\mu}\phi\nabla_{\nu}\phi-g_{\mu\nu}P.
\label{EMT}
\end{equation}
It is well known (see e.g. \cite{ArmendarizPicon:1999rj}), that
for the timelike derivatives $X>0$, the models under consideration
can be described in a hydrodynamical language by introducing an effective
four velocity %
\footnote{Note that even for $X>0$ the effective four velocity introduced in
(\ref{4Velocity}) is not necessarily future directed. However, the
analogy with the perfect fluid can be made exact by multiplying this
expression (\ref{4Velocity}) with $\pm1$ so that $u^{0}>0$. Furthermore,
it is convenient to use the analytic definition of the square root
so that every time when $\dot{\phi}$ changes its sign the square
root will change the sign as well preserving the future direction
of $u^{\mu}$ . %
} 
\begin{equation}
u_{\mu}=\frac{\nabla_{\mu}\phi}{\sqrt{2X}},
\label{4Velocity}
\end{equation}
along with the pressure
\begin{equation}
p=P\left(\phi,X\right),
\label{Pressure}
\end{equation}
the energy density
\begin{equation}
\varepsilon\left(\phi,X\right)=2XP_{,X}-P,
\label{EnergyDensity}
\end{equation}
and the sound speed %
\footnote{This formula for the sound speed was introduced for the cosmological
perturbations in \cite{Garriga:1999vw}. One can show \cite{Babichev:2007dw}
that the same expression is valid in the general case of backgrounds
with timelike field derivatives: $X>0$. %
} 
\begin{equation}
c_{s}^{2}\left(\phi,X\right)=\left(1+2X\frac{P_{,XX}}{P_{,X}}\right)^{-1}=\left(\frac{\partial p}{\partial\varepsilon}\right)_{\phi}.
\label{Sound speed}
\end{equation}
In these variables the energy momentum tensor has the form corresponding
to the one of a perfect fluid 
\[
T_{\mu\nu}=\left(\varepsilon+p\right)u_{\mu}u_{\nu}-pg_{\mu\nu}.
\]
It is convenient to use the hydrodynamical notation for these functions
of $\phi$ and $X$ also for $X\leq0$ when they do not have their
usual physical meaning of velocity etc.

\subsection{Field redefinitions and conditions for stationarity}
If the field $\phi$ does not have any direct interactions except with
gravity, then obviously a field redefinition $\phi=\phi(\tilde{\phi})$
cannot affect any observables besides the field itself. This is
a particular case of a stronger statement (see e.g. Ref. \cite{Weinberg}).
Obviously the solutions $\phi\left(x\right)$ and $\tilde{\phi}\left(x\right)$
result through Einstein equations in the same gravitational metric
$g_{\mu\nu}\left(x\right)$ and describe in that sense the same physical
process. Thus it is interesting to investigate the properties of k-\emph{Essence}
under field redefinitions. Under field redefinitions $\phi=\phi(\tilde{\phi})$
we have $\nabla_{\mu}\phi=\left(d\phi/d\tilde{\phi}\right)\nabla_{\mu}\tilde{\phi}$
whereas the expressions for the energy momentum tensor $T_{\mu\nu}$
and all hydrodynamical quantities $\varepsilon$, $p$, $c_{s}$ and
$u^{\mu}$ remain unchanged or covariant \footnote{Note that the four velocity (\ref{4Velocity}) is invariant up to the sign only.}. Here we should distinguish between covariance and invariance. Covariance
means that the way how the quantities / equations are constructed
from other objects remains unchanged whereas invariance implies exactly
the same functional dependence on these objects. For example, the formula
(\ref{EnergyDensity}) defining the energy density $\varepsilon$ through
Lagrangian $P$, $X$ and the derivative $P_{,X}$ looks the same
after a field redefinition (covariant), however the dependence of
the Lagrangian on the field does change (not invariant). It is obvious that
e.g. the value of physical energy density at every point should not
change under field redefinitions, but here these quantities reveal
in addition such covariance with respect to the field redefinitions
as it is the case for e.g. Euler-Lagrange equations. However, this
covariance is not guarantied for all interesting objects. It is worthwhile
mentioning that, e.g. the metric \cite{Babichev:2007dw}
\[
G_{\mu\nu}\left[\phi_{0}\right]=\left(\frac{P_{,X}}{c_{s}}\right)\left(g_{\mu\nu}-c_{s}^{2}\left(\frac{P_{,XX}}{P_{,X}}\right)\nabla_{\mu}\phi_{0}\nabla_{\nu}\phi_{0}\right),
\]
describing the propagation of small perturbations $\pi$ around a
given background $\phi_{0}\left(x\right)$ transforms conformally
under field redefinitions $\phi=\phi(\tilde{\phi})$: 
\[
G_{\mu\nu}\left[\phi_{0}\right]=\left(\frac{d\tilde{\phi}}{d\phi}\right)_{0}^{2}G_{\mu\nu}[\tilde{\phi}_{0}].
\]
Thus, as expected, the causal structure does not change under
field redefinitions. The conformal factor $\left(d\tilde{\phi}/d\phi\right)_{0}^{2}$
compensates for the redefinition of perturbations $\pi=\left(d\phi/d\tilde{\phi}\right)_{0}\tilde{\pi}$.

Let us further consider a stationary space-time with metric $g_{\mu\nu}$
and a timelike Killing vector $t^{\alpha}$. Thus $\pounds_{t}g_{\mu\nu}=0$,
where $\pounds_{t}$ is the Lie derivative. The configuration is stationary,
if per definition 
\[
\pounds_{t}T_{\mu\nu}=0.
\]
Using Leibniz rule we have 
\begin{eqnarray}
 &  & \pounds_{t}T_{\mu\nu}=\left(\pounds_{t}P_{,X}\right)\nabla_{\mu}\phi\nabla_{\nu}\phi-g_{\mu\nu}\pounds_{t}P+\nonumber \\
 &  & +P_{,X}\left[\left(\pounds_{t}\nabla_{\mu}\phi\right)\nabla_{\nu}\phi+\left(\pounds_{t}\nabla_{\nu}\phi\right)\nabla_{\mu}\phi\right]=0
\label{StatExpand}
\end{eqnarray}
By multiplying this expression with $g^{\mu\nu}$ we obtain 
\begin{equation}
0=\pounds_{t}T_{\mu}^{\mu}=\pounds_{t}\left(2XP_{,X}-4P\right)=\pounds_{t}\left(\varepsilon-3p\right).\label{TraceInv}
\end{equation}

Suppose the configuration $\phi\left(x^{\mu}\right)$ is such that
$\nabla_{\mu}\phi$ is a null vector: $X=0$. In that case we
can multiply the right hand side of the Eq. (\ref{StatExpand}) with
$g^{\mu\nu}$ to obtain $\pounds_{t}P=0$. Further we have $0=\pounds_{t}P=P_{,\phi}\partial_{t}\phi$.
As we are looking for stationary but not static solutions we have
$P_{,\phi}=0$. Thus the Lagrangian should be symmetric with respect
to field shifts $\phi\rightarrow\phi+c$, where $c$ is an arbitrary
constant. 

For $X\neq0$ it is convenient to introduce the projector 
\begin{equation}
\mathcal{P}_{\mu\nu}=g_{\mu\nu}-\frac{\nabla_{\mu}\phi\nabla_{\nu}\phi}{2X},
\label{Projector}
\end{equation}
with the properties 
\begin{eqnarray}
\mathcal{P}_{\mu\nu}\nabla^{\nu}\phi=0, & \mathcal{P}_{\mu\lambda}\mathcal{P}^{\lambda\nu}=\mathcal{P}_{\mu}^{\:\nu}\,\mbox{ and} & \mathcal{P}_{\mu}^{\mu}=3.
\label{ProjectorProperties}
\end{eqnarray}
Moreover, this projector is both invariant and covariant under field
reparametrizations: $\mathcal{P}_{\mu\nu}[\phi]=\mathcal{P}_{\mu\nu}[\tilde{\phi}]$.
By acting with the projector $\mathcal{P}^{\mu\nu}$ on the left hand
side of Eq. (\ref{StatExpand}) $ $we have $0=\mathcal{P}^{\mu\nu}\pounds_{t}T_{\mu\nu}=-3\pounds_{t}P$. Therefore, if the configuration is stationary then in particular 
\begin{equation}
\pounds_{t}P=0,
\label{ConstPressure}
\end{equation}
which for the hydrodynamical case reduces to the constancy of pressure
$p$. Combining this with (\ref{TraceInv}) we obtain the time independence
of the energy density $\varepsilon$ or 
\begin{equation}
\pounds_{t}\left(XP_{,X}\right)=0.
\label{ConstEnergy}
\end{equation}
Further we can act on the left hand side of Eq. (\ref{StatExpand})
with $ $$\mathcal{P}^{\alpha\nu}$: so that $0=\mathcal{P}^{\alpha\nu}\pounds_{t}T_{\mu\nu}=P_{,X}\mathcal{P}^{\alpha\nu}\left(\pounds_{t}\nabla_{\nu}\phi\right)\nabla_{\mu}\phi$. Thus, stationarity implies 
\[
\mathcal{P}^{\alpha\beta}\left(\pounds_{t}\nabla_{\beta}\phi\right)=0.
\]
Using the properties of the projector (\ref{ProjectorProperties}),
Leibniz rule and that $t^{\alpha}$ is a Killing vector one obtains
\[
0=\mathcal{P}^{\alpha\beta}\left(\pounds_{t}\nabla_{\beta}\phi\right)=-\nabla_{\beta}\phi\pounds_{t}\mathcal{P}^{\alpha\beta}=-\pounds_{t}\nabla^{\alpha}\phi+\frac{\nabla^{\alpha}\phi}{2X}\pounds_{t}X.
\]
The last expression in turn can be written in the following form 
\[
-\pounds_{t}\nabla^{\alpha}\phi+\frac{\nabla^{\alpha}\phi}{2X}\pounds_{t}X=\sqrt{2X}\pounds_{t}\left(\frac{\nabla^{\alpha}\phi}{\sqrt{2X}}\right).
\]
Therefore, stationarity implies 
\begin{equation}
\pounds_{t}\left(\frac{\nabla^{\alpha}\phi}{\sqrt{2X}}\right)=0,
\label{statVelocity}
\end{equation}
or in the hydrodynamical notation $\pounds_{t}u^{\mu}=0$ %
\footnote{The vector $u^{\mu}$ is formally imaginary for $X<0$. However, without
any change of the results one could redefine $u^{\mu}$ in this case: $u^{\mu}=\nabla^{\mu}\phi/\sqrt{-2X}$
.%
}. Thus we have proved that for any stationary configuration the following
conditions 
\begin{eqnarray}
\pounds_{t}u^{\mu}=0, & \pounds_{t}\varepsilon=0\:\mbox{ and} & \pounds_{t}p=0,
\label{StationarityCOND}
\end{eqnarray}
should be satisfied. Note that these conditions are covariant under
field redefinitions and, for the hydrodynamical case ($X>0$), are
intuitively clear requirements. Sometimes (see e.g. \cite{Frolov:2004vm})
one claims that the stationarity implies a stronger requirement:
\begin{equation}
\pounds_{t}\nabla_{\mu}\phi=0,
\label{FrolovCondition}
\end{equation}
instead of the condition (\ref{statVelocity}). However, the equation
above is not covariant under the field redefinitions and does not
follow from the stationarity of the energy momentum tensor. 

Now let us find what type of theories $P\left(\phi,X\right)$ and
field configurations $\phi\left(x^{\mu}\right)$ can, in principle,
satisfy conditions (\ref{StationarityCOND}). It is convenient to
chose a coordinate system $\left(t,x^{i}\right)$ such that the time
coordinate corresponds to the integral curves of $t^{\alpha}$. In
that case the Lie derivative reduces to the partial derivative $\pounds_{t}=\partial_{t}$.

\subsection{Which filed configurations can have constant effective four velocity
$u^{\mu}$ ?}
Now let us find the configurations $\phi\left(x^{\mu}\right)$ satisfying
the condition on the effective four velocity (\ref{statVelocity}). For
the time component of the four velocity we have 
\begin{equation}
\partial_{t}\left(\frac{\dot{\phi}}{\sqrt{2X}}\right)=\frac{\ddot{\phi}}{\sqrt{2X}}+\dot{\phi}\partial_{t}\left(\frac{1}{\sqrt{2X}}\right)=0,
\label{VelTcompCond}
\end{equation}
where $\dot{\phi}=\partial_{t}\phi$, while for the spatial components
\begin{equation}
\partial_{t}\left(\frac{\partial_{i}\phi}{\sqrt{2X}}\right)=\frac{\partial_{i}\dot{\phi}}{\sqrt{2X}}+\partial_{i}\phi\partial_{t}\left(\frac{1}{\sqrt{2X}}\right)=0.
\label{VelXcompCond}
\end{equation}
Obviously these equations have a trivial static solution $\phi=\phi\left(x^{i}\right)$.
To find a nontrivial solution we combine these two equations to obtain
following system of equations 
\[
\dot{\phi}\partial_{i}\dot{\phi}-\ddot{\phi}\partial_{i}\phi=0,
\]
which is equivalent to 
\begin{equation}
\partial_{t}\left(\frac{\partial_{i}\phi}{\dot{\phi}}\right)=0.
\label{secondOrder}
\end{equation}
This is a system of partial differential equations of the second order.
Integrating Eq. (\ref{secondOrder}) we obtain the following linear
homogeneous system 
\begin{equation}
\partial_{i}\phi=V_{i}\left(x^{j}\right)\dot{\phi},
\label{LinearSyst}
\end{equation}
where $V_{i}\left(x^{j}\right)$ are unknown time independent functions.
This is the first order system of three partial differential equations
for only one function $\phi$. Let us find the consistency conditions
under which the system can have solutions. Differentiating $i-$equation
with respect to $x^{j}$ and using the time differentiation of the
$j-$equation we obtain 
\[
\partial_{j}\partial_{i}\phi=\partial_{j}V_{i}\dot{\phi}+V_{i}\partial_{j}\dot{\phi}=\partial_{j}V_{i}\dot{\phi}+V_{i}V_{j}\ddot{\phi}.
\]
Now we can compare this result with the result of the same procedure
performed for the $j-$equation. We obtain 
\[
\partial_{i}V_{j}-\partial_{j}V_{i}=0.
\]
For a simply connected manifold, the last equation implies the existence
of a function (potential) $\Psi\left(x^{i}\right)$ such that $V_{i}=\partial_{i}\Psi$.
Otherwise there are no solutions for (\ref{LinearSyst}). 

For the $i-$equation we can assume that all $x^{k}$ with $k\neq i$
are frozen parameters and for the characteristics (for the method of characteristics
see e.g. excellent book \cite{Arnold}) we obtain 
\begin{eqnarray*}
\frac{dt}{d\tau}=-\partial_{i}\Psi\left(x^{j}\right), &  & \frac{dx^{i}}{d\tau}=1.
\end{eqnarray*}
The first integral $\mathcal{I}$ of this system is given by the
constant of integration for the equation 
\[
\frac{dt}{dx^{i}}=-\partial_{i}\Psi\left(x^{j}\right).
\]
By integrating which we obtain 
\[
t=\mathcal{I}-\Psi\left(x^{i}\right),
\]
therefore the general solution $\phi\left(t,x^{i}\right)$ is given
as an arbitrary function of the first integral $\mathcal{I}$ $ $:
\begin{equation}
\phi\left(t,x^{i}\right)=\Phi\left(t+\Psi\left(x^{i}\right)\right).
\label{general solution}
\end{equation}
Thus the general solution for equations (\ref{VelTcompCond}) and
(\ref{VelXcompCond}) contains two arbitrary functions. Note that
the system (\ref{secondOrder}) does not have any other general solutions
besides (\ref{general solution}). It is easy to prove that this
solution satisfies the equations (\ref{VelTcompCond}) and (\ref{VelXcompCond}).
Indeed we have 
\begin{eqnarray*}
\dot{\phi}=\frac{d\Phi}{d\mathcal{I}} & \mbox{and} & \partial_{i}\phi=\frac{d\Phi}{d\mathcal{I}}\partial_{i}\Psi,\end{eqnarray*}
therefore 
\begin{equation}
X=\frac{1}{2}\left(\frac{d\Phi}{d\mathcal{I}}\right)^{2}\left(g^{00}+2g^{0i}\partial_{i}\Psi+g^{ik}\partial_{i}\partial_{k}\Psi\right),
\label{X}
\end{equation}
and the time component 
\[
\frac{\dot{\phi}}{\sqrt{2X}}=\frac{1}{\sqrt{g^{00}+2g^{0i}\partial_{i}\Psi+g^{ik}\partial_{i}\partial_{k}\Psi}},
\]
along with the spatial components
\[
\frac{\partial_{i}\phi}{\sqrt{2X}}=\frac{\partial_{i}\Psi}{\sqrt{g^{00}+2g^{0i}\partial_{i}\Psi+g^{ik}\partial_{i}\partial_{k}\Psi}},
\]
are obviously time independent because the metric is stationary.
It is worth mentioning that by using the condition (\ref{FrolovCondition})
we would arrive at the general solution $\phi\left(t,x^{i}\right)=t+\Psi\left(x^{i}\right)$,
missing the arbitrary functional dependence $\Phi$. Note that arbitrary
field redefinitions correspond to the freedom in choosing $\Phi$.

\subsection{Which Lagrangians do allow for the stationary configurations?}
Now let us consider the restrictions on $P\left(\phi,X\right)$ arising
from the requirement that the pressure and energy density should be
time independent for the general solution (\ref{general solution}).
From Eq. (\ref{ConstPressure}) we have 
\begin{equation}
\partial_{t}P=P_{,\phi}\dot{\phi}+P_{,X}\dot{X}=0,
\label{difP}
\end{equation}
while from Eq. (\ref{ConstEnergy}) 
\[
\partial_{t}\left(XP_{,X}\right)=\dot{X}P_{,X}+XP_{,X\phi}\dot{\phi}+XP_{,XX}\dot{X}=0.
\]
Eliminating $\dot{X}$ from these equations results in 
\begin{equation}
XP_{,X\phi}-\left(XP_{,XX}+P_{,X}\right)\frac{P_{,\phi}}{P_{,X}}=0.
\label{EqForLagrangian}
\end{equation}
This equation is a second order partial differential equation for
$P\left(\phi,X\right)$. A trivial solution of this equation is a
shift symmetric Lagrangian $P\left(X\right)$. It is well known that
shift symmetric theories are exactly equivalent to hydrodynamics for
$X>0$. Obviously hydrodynamics usually allows for the steady flows.
Let us find a general solution of the equation (\ref{EqForLagrangian}).
This general solution should depend on two arbitrary functions. It
is convenient to rewrite Eq. (\ref{EqForLagrangian}) in the following
form 
\[
\frac{\partial\ln\left(P_{,\phi}/P_{,X}\right)}{\partial\ln X}=1.
\]
Integrating this equation we obtain 
\begin{equation}
P_{,\phi}=\sigma\left(\phi\right)XP_{,X},
\label{LinearEqForP}
\end{equation}
where $\sigma\left(\phi\right)$ is an arbitrary function. The last
equation (\ref{DynSys}) is a linear partial differential equation
of the first order. Similarly to our previous calculations we use
the method of characteristics to find the general solution. For the
characteristics we have 
\begin{eqnarray}
\frac{d\phi}{d\tau}=1 & \mbox{and} & \frac{dX}{d\tau}=-\sigma\left(\phi\right)X,
\label{DynSys}
\end{eqnarray}
thus the integral curves are given by the equation 
\[
\frac{dX}{d\phi}=-\sigma\left(\phi\right)X.
\]
The general solution of the last equation is 
\[
X=\mathfrak{I}\exp\left(-\int\sigma\left(\phi\right)d\phi\right),
\]
where $\mathfrak{I}$ is a constant of integration. Thus the general
solution to the equations (\ref{LinearEqForP}) and (\ref{EqForLagrangian})
is an arbitrary function of the first integral $\mathfrak{I}$$ $
of the dynamical system (\ref{DynSys}): 
\begin{equation}
P\left(\phi,X\right)=F\left(X\mbox{e}^{f\left(\phi\right)}\right),
\label{GenLagrangian}
\end{equation}
where $F$ and $f\left(\phi\right)=\int\sigma\left(\phi\right)d\phi$
are arbitrary functions. Note that all solutions of (\ref{EqForLagrangian})
are described by (\ref{GenLagrangian}). It is obvious that the Lagrangian
(\ref{GenLagrangian}) has a hidden shift symmetry. Namely, we can
always perform a field redefinition 
\begin{equation}
\tilde{\phi}\left(\phi\right)=\int d\phi\:\mbox{e}^{f\left(\phi\right)/2},
\label{neededFieldRedef}
\end{equation}
so that the new Lagrangian is shift symmetric $P\left(\phi,X\right)=F(\tilde{X})$,
where $\tilde{X}=\frac{1}{2}g^{\mu\nu}\nabla_{\mu}\tilde{\phi}\nabla_{\nu}\tilde{\phi}=X\mbox{e}^{f\left(\phi\right)}$. Thus all scalar field theories which allow for stationary configurations
are necessarily shift symmetric (explicitly or after field redefinition).
Further, we will use the notation $\tilde{\phi}$ always for such field
variables in which the system is invariant under shift transformations
$\tilde{\phi}\rightarrow\tilde{\phi}+c$, where $c$ is an arbitrary
constant . 

Finally we can specify the profiles $\Phi$ of stationary configurations.
Equations (\ref{difP}) and (\ref{X}) yield 
\[
P_{,\phi}+P_{,X}\left(\frac{d^{2}\Phi}{d\mathcal{I}^{2}}\right)\left(g^{00}+2g^{0i}\partial_{i}\Psi+g^{ik}\partial_{i}\partial_{k}\Psi\right)=0,
\]
and using Eq. (\ref{LinearEqForP}) and (\ref{X}) we obtain 
\begin{equation}
\frac{d^{2}\Phi}{d\mathcal{I}^{2}}+\frac{1}{2}\left(\frac{d\Phi}{d\mathcal{I}}\right)^{2}\sigma\left(\Phi\right)=0.
\label{PhiEq}
\end{equation}
We know that in terms of the new field $\tilde{\phi}$ the Lagrangian
is shift symmetric. Thus for this parametrization $\sigma(\tilde{\phi})=0$. Therefore $\tilde{\Phi}\left(\mathcal{I}\right)=\alpha\mathcal{I}+\beta=t+\Psi\left(x^{i}\right)$
where we have absorbed the constants into $\Psi$ and $t$. Thus in
terms of the field variable $\tilde{\phi}$, in which the theory
is shift symmetric, the possible stationary configurations are always
given by 
\begin{equation}
\tilde{\phi}=t+\Psi\left(x^{i}\right),
\label{RES}
\end{equation}
and we are back to the usual ansatz (\ref{FrolovCondition}). The
stationary configurations in terms of the field variable $\phi$
can be obtained by solving equation (\ref{neededFieldRedef}) or (\ref{PhiEq})
with respect to $\phi$. This procedure determines the function $\Phi$.
While the function $\Psi\left(x^{i}\right)$ has to be fixed from
the equations of motion and boundary/initial conditions. 

It is worth noting that if the metric $g_{\mu\nu}$ possesses
another Killing vector corresponding to e.g. axial symmetry $ $$\pounds_{\theta}g_{\mu\nu}=0$
then we can apply the result (\ref{RES}) to the angular variable $\theta$.
Thus the solution is 
\[
\tilde{\phi}=t+\Omega\theta+\Psi\left(x_{\perp}^{i}\right),
\]
where $\Omega$ is a constant and $x_{\perp}^{i}$ denotes the rest
of the coordinates.

\section{Conclusions and Discussion}
In this paper we have proved that the existence of stationary configurations
requires shift symmetry. Namely (may be after a field redefinition)
the system has to be invariant with respect to the transformation $\tilde{\phi}\rightarrow\tilde{\phi}+c$,
for all constants $c$. The result is valid in the self-consistent
case where the geometry is produced by the scalar field as well as
in the test field approximation where the stationary field configuration
appears on the gravitational background governed by other sources.
The shift symmetry implies the conservation of the Noether current
\[
J_{\mu}=P_{,\tilde{X}}\nabla_{\mu}\tilde{\phi}.
\]
Interestingly, the equation of motion implies $\nabla_{\mu}J^{\mu}=0$, which is a statement of the conservation of the current $J^{\mu}$. In the case when $\nabla_{\mu}\tilde{\phi}$ is timelike the current $J^{\mu}$ can be written in the form of an effective particle density current $J^{\mu}=\tilde{n}u^{\mu}$,
where the particle density \footnote{Note that this number density is none other than the canonical momenta for the field $\tilde{\phi}$ in the co-moving reference frame.} is 
\[
\tilde{n}=\sqrt{2\tilde{X}}P_{\tilde{,X}}.
\]
Note that this current is not covariant under field redefinitions.
The conservation $\nabla_{\mu}J^{\mu}=0$ of the particle density
current usually holds in the standard hydrodynamics. However, the most
interesting models of cosmological scalar fields do not posses this
additional conservation law associated with the shift symmetry. Thus
the result obtained in this paper implies that there is no exact Bondi
(steady flow) accretion for popular classes of models for dynamical
\emph{Dark Energy} like \emph{Quintessence} and k-\emph{Essence.}
This\emph{ }result may not have a very strong qualitative impact on
the growth of Black Holes or on the evolution of the cosmological fields
around them. Indeed, one should expect that the accretion rate should
be in any case rather small (for the case of canonical scalars see
Ref. \cite{Frolov:2002va}) . Especially in the late / current universe,
one can almost always neglect the growth of the Black Hole along with
the corresponding backreaction. Nevertheless, this result changes
the setup for the investigation of the problem. Now in order to study
how these fields could accrete onto Black Holes one is forced to solve
the Cauchy problem for nonlinear partial differential equations, instead
of solving the boundary problem for nonlinear ordinary differential
equations. In particular to approach this problem one has to choose
some initial configuration for the field and it's time derivative. At this
stage, it is not clear what are reasonable, physically motivated
initial conditions and at what time they should be posed. This is
very different from the case of Bondi accretion where the boundary
conditions are fixed by cosmological evolution and the membrane property
of the BH horizon. However, it may happen that there are some special
attractor or self-similar regimes to which the solutions would approach
in the late time asymptotic. Nevertheless, one cannot guarantee either
the existence of these attractors nor their uniqueness for a general
model. Moreover, even if a unique attractor exists, then it is not
a priory known how wide the base of attraction is in the phase space
consisting of initial configurations of the field and its time derivative.
Thus, the procedure for finding these attractor solutions is not only
a predominantly numerical exercise, but also generically not very promising and predictive.
Nevertheless, it is very interesting to find examples of scalar field
systems possessing solutions of this type. In \cite{Frolov:2002va}
it was demonstrated that for canonical scalars and many potentials
the solutions indeed approach the steady flow. 

In addition one has to mention that having a shift symmetric theory
is a necessary, but not sufficient for the existence of a stationary
configuration. For example, in hydrodynamics there can be either exceptional
theories or even exceptional boundary conditions for which there are
no stationary configurations. In particular the simple accretion of
dust onto a Black Hole occurs along geodesics and therefore is not
steady. A similar situation happens in the case of the\emph{ Ghost
Condensate} for which the accretion rate blows up when the field configuration
at spatial infinity approaches the condensation point (compare Refs. \cite{Frolov:2004vm} and \cite{Mukohyama:2005rw}). Moreover, in
the DBI model considered in \cite{Babichev:2006vx,Babichev:2007wg} it was found that
a physically meaningful steady state accretion is not possible when
the sound speed at spatial infinity is $c_{s}^{2}>4/3$. 

In this paper we have considered only a single self interacting scalar
field. It would be interesting to study other types of fields, in
particular one could think of scalars with internal degrees of freedom
e.g charged scalars accreting onto a charged Black Hole. We expect
that the appearance of new external forces and internal degrees of
freedom can change the picture. Another interesting problem is to
find possible attractor or self-similar asymptotic solutions and develop
a perturbation theory around them. As we have shown stationary configurations
are possible only for theories which are equivalent to perfect fluids.
This result reveals once again that the relation between hydrodynamics
and field theory is rather deep. Therefore we think this connection
deserves a further study. We found that investigation of possible dynamical
backgrounds around Black Holes is interesting not only from the point
of view of mathematical physics but may be relevant for a better understanding
of both Black Holes physics and may be even the nature of \emph{Dark
Energy}. 

\begin{acknowledgments}
It is a pleasure to thank Andrei Gruzinov, Andrew MacFadyen, Oriol
Pujolas, Ignacy Sawicki and Eugeny Babichev for interesting valuable
discussions. The work of AV was supported by the James Arthur Fellowship.
AV would like to thank the Michigan Center for Theoretical Physics
for the kind hospitality during the preparation of this paper.
\end{acknowledgments}
\bibliographystyle{JHEP}
\bibliography{Base}

\end{document}